\documentclass[prb,showpacs,superscriptaddress,twocolumn,floats,citeautoscript]{revtex4}

\usepackage{graphicx}
\usepackage{amsmath}
\usepackage{amssymb}
\usepackage{bm}
\usepackage{times}
\usepackage{colordvi}

\graphicspath{{.}{./EPS/}}

\newcommand* {\ee}{\ensuremath{\mathrm{e}}}
\newcommand* {\vek}[1]{{\ensuremath{\bm{\mathrm{#1}}}}}

\begin{document}

\title{Large variations in the hole spin splitting of quantum-wire subband edges} 

\author{D. Csontos}
\affiliation{Institute of Fundamental Sciences, Massey University, Private Bag
11~222, Palmerston North, New Zealand}

\author{U. Z\"ulicke}
\affiliation{Institute of Fundamental Sciences, Massey University, Private Bag
11~222, Palmerston North, New Zealand}
\affiliation{MacDiarmid Institute for Advanced Materials and Nanotechnology,
Massey University, Palmerston North, New Zealand}

\date{\today}

\begin{abstract}

We study Zeeman splitting of zone-center subband edges in a cylindrical hole wire
subject to a magnetic field parallel to its axis. The $g$-factor turns out to fluctuate
strongly as a function of wire-subband index, assuming values that differ substantially
from those found in higher-dimensional systems. We analyze the spin properties
of hole-wire states using invariants of the spin-3/2 density matrix and find a strong
correlation between $g$-factor value and the profile of hole-spin polarization density.
Our results suggest possibilities for confinement engineering of hole spin splittings.

\end{abstract}


\maketitle

Spin splitting of charge carriers in semiconductors has been a focus of recent
research interest, partly because it may form the basis for the new paradigms
of spin-based electronics and quantum information processing~\cite{reviews}.
Besides such possible applications, intriguing fundamental-science questions
motivate the study of charge carriers' spin properties. In particular, the
quantum-mechanical coupling between spin and orbital degrees of freedom
enables a host of, sometimes counterintuitive, mechanisms for manipulating
spins in nanostructures~\cite{crook:prl:05}. As states in the valence band of a
typical semiconductor are subject to a strong spin-orbit coupling, hole spin
splittings will be highly tunable by geometrical and quantum-confinement effects.
Large anisotropies of hole $g$-factors in quantum wells~\cite{rolandbook}, point
contacts~\cite{uz:prl:06}, quantum dots~\cite{pryor:prl:06}, and localized
acceptor states~\cite{haug:prl:06} provide pertinent examples. The origin of
such peculiar hole-spin properties can be traced to the fact that quasiparticles
from the top-most valence band are characterized by total angular momentum
(spin) 3/2. (Conduction-band electrons are spin-1/2 particles like electrons in
vacuum.) Although spin is an intrinsically quantum degree of freedom, it has
been possible to rationalize spin-1/2 physics largely in terms of a
magnetic-dipole analogy. This classical-physics-inspired vehicle for our
understanding succeeds because any spin-1/2 density matrix is fully
characterized~\cite{dftBook} by the (trivial) particle density and a dipole
moment associated with spin-polarization. In contrast, the spin-3/2 density
matrix has two more invariants, a quadrupole and an octupole
moment~\cite{roland:prb:04}. As a surprising implication, magnetic fields
do not always induce a spin polarization in two-dimensional (2D) hole
systems~\cite{roland:prb:05a}.

\begin{figure}[b]
\includegraphics[width=3.2in]{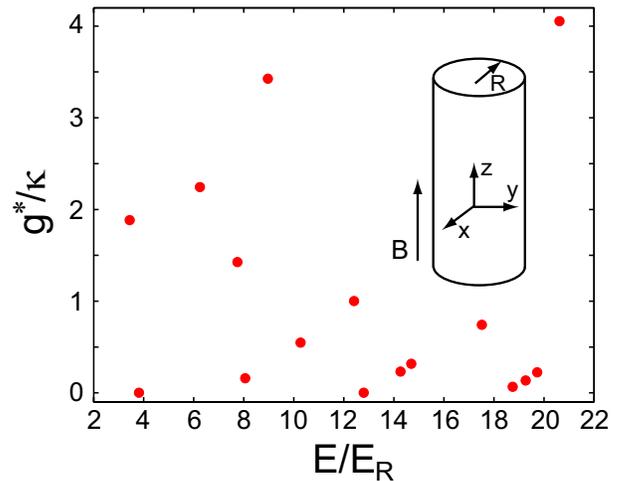}
\caption{(Color online) Land\'e factors $g^\ast$ of low-lying subband edges in
a cylindrical hole quantum wire (in units of the bulk hole $g$-factor $\kappa$).
The abscissa shows subband energies measured from the valence-band top
in units of $E_R=-\gamma_1 \hbar^2/(2 m_0 R^2)$. $\gamma_1$ is a
materials-dependent (Luttinger) parameter, $m_0$ the electron mass in
vacuum, and $R$ the wire radius. The calculation is based on the spherical
Luttinger model~\cite{luttham2} for the top valence band and assumed
the GaAs value (0.37) for the relative spin-orbit coupling strength
$\gamma_{\text{s}}/\gamma_1$.\label{fig:1}}
\end{figure}
Our theoretical study presented here reveals the drastic influence of a
quantum-wire confinement on spin-3/2 physics. Figure~\ref{fig:1} shows
universal results for the Land\'e $g$-factors of low-lying subband edges in
cylindrical quantum wires subject to a parallel magnetic field. Two surprising
features are apparent. First, $g^\ast$ is seen to assume unusual values. 
Considering the wire axis to be a natural quantization axis for hole
spin and remembering that spin-3/2 projection eigenvalues are $\pm 3/2$
and $\pm 1/2$, we would expect to find only $6\kappa$ and $2\kappa$ as
possible $g$-factors. (Here $\kappa$ is the hole $g$-factor in the bulk
material~\cite{luttham2}.) In Fig.~\ref{fig:1}, quite different numbers are
found. Second, a strong variation of $g$-factor values between different
subbands is observed. These features contrast also with situations
found in 2D hole systems where confinement  results in the
suppression of field-induced spin polarization~\cite{rolandbook} and
$g$-factors close to $0$ and $4\kappa$  emerge~\cite{hensel:prb:74,
OldHoleZ1,OldHoleZ2}.

In the following, we show how the anomalous behavior of wire-subband spin
splittings is rooted in peculiar patterns of hole-spin dipole density that arise from
the interplay between quasi-onedimensional (1D) confinement and spin-orbit
coupling in the valence band. Our results are experimentally testable, as
Zeeman splitting of quasi-1D subband edges can be measured directly in
transport experiments~\cite{patel:prb:91}. P-type versions of recently fabricated
highly symmetric semiconductor nanowires~\cite{nanoWireRev1} 
should be particularly well-suited for observing effects discussed in this work. 
Our study is also relevant for understanding spin properties of quasi-1D 
hole systems realized in 2D semiconductor heterostructures~\cite{piccio:apl:05,
romain:apl:06,uz:prl:06} and provides physical insight for the interpretation of
recent numerical results~\cite{kossut:prb:00,xia:epjb:06,kita:prb:06}.

The remainder of this article is organized as follows. We first describe the
theoretical formalism for calculating hole-wire subband energies and
$g$-factors. After that, the characterization of spin-3/2 density matrices in
terms of multipole invariants is introduced and results for the cylindrical-wire
subband edges are presented. The emerging correlation between hole
spin-dipole density and $g$-factor will be discussed. In closing, we comment
on the effect of deviations from certain idealities assumed in our theoretical
description.

\textit{Cylindrical hole quantum wire in a parallel magnetic field\/}.
We adopt the Luttinger model~\cite{luttham2} to describe states in the
top-most valence band of a bulk semiconductor. The corresponding
Hamiltonian operates in the $4\times 4$ space spanned by eigenstates of
spin-3/2 projection on an arbitrary axis with quantum numbers $\pm 3/2$
(heavy holes, HH) and $\pm 1/2$ (light holes, LH). It reads in the spherical
approximation~\cite{luttham2,lip:prl:70}
\begin{equation}\label{eq:luttham}
{\mathcal H}_{\text{L}} = - \frac{\gamma_1}{2 m_0} \, p^2+
\frac{\gamma_{\text{s}}}{m_0} \left[ \left( \vek p \cdot \hat{\vek J} \right)^2
- \frac{5}{4} \, p^2 \,{\mathbf 1}_{4\times 4} \right] \quad .
\end{equation}
We denote linear orbital hole momentum by $\vek p$, $\hat{\vek J}$ is
the vector of spin-3/2 matrixes, and $m_0$ the vacuum electron mass. 
Energies are measured from the valence-band edge. The ratio $\bar\gamma
\equiv \gamma_{\text{s}}/\gamma_1$ of Luttinger band-structure parameters
quantifies the relative strength of spin-orbit coupling in the valence band.

We consider a wire that is parallel to the $z$ axis and subject to Zeeman
splitting due to a magnetic field pointing in the same direction. Thus it is
useful to choose the representation where $\hat J_z$ is diagonal. As we are
focused on finding hole-wire subband edges at the zone center~\cite{offCent},
we set $p_z=0$. Orbital magnetic-field effects are neglected~\cite{DMSappl}.
To  determine bulk hole states that can be superimposed to find cylindrical-wire
subband states, we use polar coordinates $r$, $\varphi$ and the wave-function
\textit{ansatz\/}~\cite{sercel:prb:90}
\begin{equation}\label{eq:Ansatz}
\psi(r, \varphi) = \ee^{i m \varphi} \, \left( \begin{array}{c} a_m J_{m}(k r) \\
\ee^{i \varphi} b_m J_{m+1} (k r) \\\ee^{2 i \varphi} c_m J_{m+2}(k r) \\
\ee^{3i \varphi} d_m J_{m+3}(k r) \end{array} \right) \quad .
\end{equation}
$J_m(x)$ is a Bessel function, and $a_m,\dots, d_m$ are constants.
Uniqueness of the hole wave function requires $m$ to  be an integer. Inserting
Eq.~(\ref{eq:Ansatz}) into the stationary Schr\"odinger equation simplifies the
Hamiltonian for hole motion transverse to the $z$ axis, which now
reads~\cite{sercel:prb:90}
\begin{equation}\label{eq:perpHam}
{\mathcal H}_{\text{L}}^{(\perp)} = - \frac{\gamma_1\hbar^2 k^2}{2 m_0} \left \{
{\mathbf 1}_{4\times 4} + \bar\gamma \left[ {\hat J}_z^2 - \frac{5}{4} {\mathbf
1}_{4\times 4} + {\hat J}_+^2 +  {\hat J}_-^2 \right] \right\} .
\end{equation}
Here $\hat J_\pm = ( \hat J_x \pm i \hat J_y)/\sqrt{2}$ are ladder operators for
the $z$-axis projection of spin.

Zeeman splitting due to a magnetic field with magnitude $B$ applied parallel
to the $z$ axis is described by ${\mathcal H}_{\text{Z}} = 2 \kappa \mu_B
B \hat J_z$, which needs to be added to Eqs.~(\ref{eq:luttham}) and
(\ref{eq:perpHam}). $\mu_B$ is the Bohr magneton, $\kappa$ the bulk hole
$g$-factor, and we neglected the small anisotropic part of Zeeman splitting in
the bulk valence band.

Diagonalization of ${\mathcal H}_{\text{L} }^{(\perp)}+{\mathcal H}_{\text{Z}}$
yields bulk-hole eigenstates that can be superimposed to find wire-subband
bound states:
\begin{subequations}\label{eq:radWF}
\begin{eqnarray}
\psi_{km\alpha +} (r) &=&  \left(a_{\alpha +} J_m(k r), 0, \, J_{m+2}(k r), 0
\right)^{\mathrm T} , \\
\psi_{km\alpha -} (r) &=&  \left(0, J_{-m-2}(k r), 0, a_{\alpha -} J_{-m}(k r)
\right)^{\mathrm T}  .
\end{eqnarray}
\end{subequations}
We omit the polar-angle ($\varphi$)-dependent part since the wire
potential is cylindrically symmetric. In Eq.~(\ref{eq:radWF}),
$\alpha=\pm 1$ labels eigenstates within subspaces spanned by HH
and LH basis states having spin projection $\pm 3/2$ and $\mp1/2$ (for
$\psi_{km\alpha\pm}$, respectively). The coefficients $a_{\alpha \sigma}$ are
given by
\begin{equation}
a_{\alpha\sigma} = \frac{1}{\sqrt{3}} \left[ 1- 2 \alpha\sqrt{1 + \zeta_{k
\sigma} + \zeta_{k\sigma}^2} + 2 \zeta_{k\sigma} \right] \, ,
\end{equation}
where we used the abbreviation $\zeta_{k\sigma} = \sigma \kappa e B /
(\gamma_{\text{s}} \hbar k^2)$, and $-e$ is the electron charge. The 
corresponding eigenenergies are, like the $a_{\alpha\sigma}$, independent of
orbital angular momentum quantum  number $m$:
\begin{equation}
E_{\alpha\sigma}(k)= -\frac{\gamma_1\hbar^2 k^2}{2 m_0} \left[ 1 - 2 \alpha
\bar\gamma \sqrt{1 + \zeta_{k\sigma} + \zeta_{k\sigma}^2} + \bar\gamma
\zeta_{k\sigma} \right] .
\end{equation}
These results generalize previous analytical treatments~\cite{sercel:prb:90} of
cylindrical hole wires to the case with finite Zeeman splitting.

\begin{figure}[b]
\includegraphics[width=3.2in]{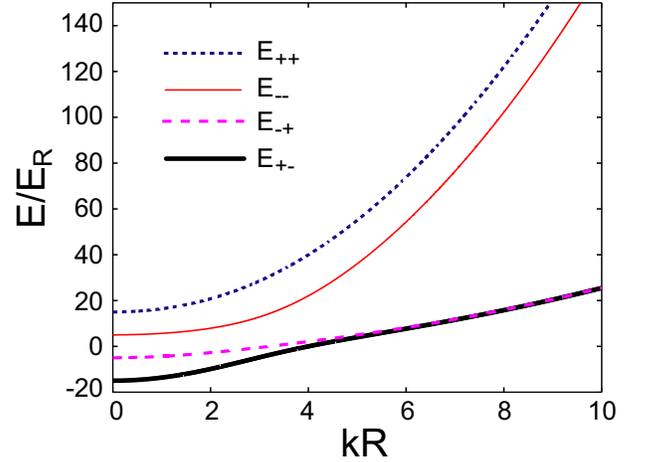}
\caption{(Color online) Energy dispersion $E_{\alpha\sigma}(k)$ of bulk-hole
states that can be superimposed to form wire-subband bound states. Here $R$
is an arbitrary length scale that will later be identified with the wire radius. The
energy unit is $E_R=-\gamma_1 \hbar^2 / (2 m_0 R^2)$, and a magnetic field
$B=10\cdot\gamma_1 \hbar / (2\kappa e R^2)$ has been applied parallel to the
$z$ direction. The GaAs value $\bar\gamma=0.37$ was assumed for the
calculation.
\label{fig:2}}
\end{figure}
Inspection of the bulk dispersions in a finite magnetic field
(Fig.~\ref{fig:2}) reveals an interesting cross-over behavior. At small $k$, i.e.,
small kinetic energy for motion perpendicular to the magnetic-field direction, a
Zeeman splitting of HH and LH states with Land\'e factor in accordance with
naive expectation ($g^\ast_{\text{HH}} = 6\kappa$ and $g^\ast_{\text{LH}} =
2\kappa$) is found. Evidently, these states have their spin closely aligned with
the $z$ axis. In contrast, at large $k$, the less-dispersive branches show no
Zeeman splitting, while that of the more dispersive ones is characterized by
$g$-factor $4\kappa$. The latter situation is reminiscent of the Zeeman effect
in 2D hole systems for an in-plane magnetic field~\cite{rolandbook}. 

Hole-wire subband bound states are found by superimposing spinors
$\psi_{k_{\alpha\sigma} m\alpha\sigma}(r)$ with fixed $m,\sigma$ and 
$E_{\alpha\sigma}(k_{\alpha\sigma}) = E$ to satisfy a hard-wall boundary
condition at $r=R$. The resulting secular equation reads
\begin{equation}\label{eq:secular}
a_{+} J_m (k_{+} R) J_{m+2}( k_{-} R) - a_ {-} J_m (k_{-} R) J_{m+2}( k_{+} R)
= 0 \, ,
\end{equation}
with conserved index $\sigma$ suppressed for the sake of brevity. Solving
Eq.~(\ref{eq:secular}) for $E$ yields the subband energies $E_{n m \sigma}
(B)$ for a cylindrical hole wire subject to a parallel magnetic field of magnitude
$B$. In general, these are mixtures of the bulk states $\psi_{km\pm\sigma}$.
Exceptions are states with $m=-1$, for which Eq.~(\ref{eq:secular})
specializes to the condition $J_{1}(k_{\pm}R)=0$ that can be satisfied by the
individual states $\psi_{km+\sigma}$ and $\psi_{km-\sigma}$. At finite $B$,
levels $E_{n m \pm}$ are Zeeman-split. We calculate the corresponding
$g$-factors using~\cite{gFactDef}
\begin{equation}\label{gStar}
g^\ast_{nm}=\lim_{B\to 0} \left| \frac{E_{n m +}(B) - E_{n m -}(B)}{\mu_{\text{B}}
B} \right| \quad .
\end{equation}
Table~\ref{tab:levels} summarizes results obtained for the ten lowest wire
subbands when the GaAs value for $\bar\gamma$ is assumed. The strong
variation of $g^\ast$ for different wire levels is illustrated in Fig.~\ref{fig:1}.
\begin{table}
\caption{\label{tab:levels}
Energies and effective $g$-factors for the lowest ten quasi-1D subband edges
in a cylindrical hole wire. Units and parameters are the same as given in the
caption of Fig.~\ref{fig:1}.}
\begin{ruledtabular}
\begin{tabular}{c|cccccccccc}
$E/E_R$ & 3.44 & 3.82 & 6.25 & 7.74 & 8.06 & 8.97 & 10.3 & 12.4 & 12.8 & 14.3 \\
$m$  & -2 & -1 & 0 & -3 & -2 & 0 & 1 & -4 & -1 & -3 \\
$g^\ast / \kappa$ & 1.88 & 0.00 & 2.23 & 1.42 & 0.161 & 3.43 & 0.545 & 1.00 & 0.00 & 0.232 
\end{tabular}
\end{ruledtabular}
\end{table}

The first subband has $g^\ast$ close to $2\kappa$, as expected for a LH
state with spin polarization parallel to the wire axis~\cite{uz:pssc:06}. The
second level belongs to the above-mentioned special class of states with
$m=-1$, which are pure bulk states and therefore have $g$-factor equal to
either $0$ or $4\kappa$. To obtain a fuller understanding of how various
effective $g$-factor values for hole-wire subbands edges emerge, we analyze
the corresponding bound states in terms of invariants for the spin-3/2 density
matrix.

\textit{Spin polarization of hole-wire bound states\/}. The peculiar physical
properties of hole quantum wires can be attributed, in a very general way, to
finite HH-LH mixing present even at the subband edges~\cite{bastardrev}. A
quantitative characterization of the latter is typically attempted by considering
expectation values for spin-3/2 projections on a fixed axis~\cite{bastardrev,
kossut:prb:00,kita:prb:06}. In particular for a wire geometry, the usefulness of
such an approach can be limited by ambiguities in the choice of a suitable
projection axis. Hence, a characterization of hole states in terms of scalar
invariants would be much more meaningful. In the following, we provide such
an analysis in terms of a multipole expansion of spin-3/2 density 
matrices~\cite{roland:prb:04}.

As hole states are four-spinors, there exist four scalars $\rho_0, \dots, \rho_3$
to characterize their spin-3/2 density matrix. Here we adopted the notation of
Ref.~\onlinecite{roland:prb:04}, where $\rho_0$ is proportional to the hole
(charge) density, $\rho_1$ is a dipole moment related to the hole-spin
polarization, the quadrupole moment $\rho_2$ quantifies HH-LH mixing, and
$\rho_3$ is an octupole moment. It is an intriguing feature of spin-3/2 physics
that magnetic fields can induce a substantial octupole moment instead of a
spin polarization (dipole moment) in 2D hole systems~\cite{roland:prb:04}. Here
we observe an analogous property of hole spin for individual states at the
quasi-1D subband edges and relate our findings to the measurable $g$-factors
of these states. 

\begin{figure}[b]
\includegraphics[width=3.2in]{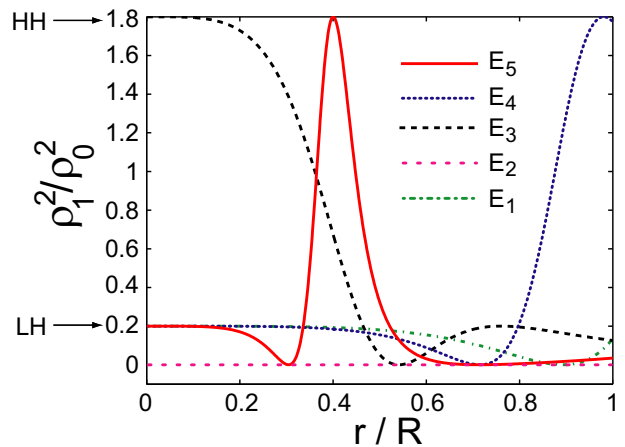}
\caption{(Color online) Radial profile of the (normalized squared magnitude of)
local hole-spin polarisation, calculated for the five lowest wire-subband-edge
bound states. Pure HH (LH) states, defined by spin projection parallel to the
$z$ axis (i.e., the wire axis), have $\rho_1^2/\rho_0^2=9/5$ ($1/5$).
\label{fig:3}}
\end{figure}

Within our model for a hole quantum wire, states at the subband edge are
superpositions of HH and LH amplitudes corresponding to $\hat J_z$ projection
$\sigma\, 3/2$ and $-\sigma\, 1/2$, respectively. [See Eqs.~(\ref{eq:radWF}).]
For these wave functions, relations exist between the multipole invariants of the
spin-3/2 density matrix:
\begin{equation}\label{eq:InvRelate}
\rho_0^2 = 2 \rho_2^2 = \rho_1^2 + \rho_3^2 \quad .
\end{equation}
The left equality in Eq.~(\ref{eq:InvRelate}) quantifies the HH-LH mixing that is
present in subband-edge states. The right equality in the same equation implies
that, for each of these states, the magnitudes of spin polarization and octupole
moment are complementary. In particular, a state with no spin polarization will
have a maximum octupole moment and \textit{vice versa\/}. We are thus able to
fully characterize the spin properties of each subband-edge bound state by
considering only $\rho_1^2/\rho_0^2$. When derived from the radial part of the
wire-subband wave function, this quantity provides a measure of local hole spin
polarization. Figure~\ref{fig:3} shows results for the five lowest cylindrical-wire
subband edges. The (mainly) LH character of the lowest subband is apparent
($\rho_1^2/\rho_0^2$ is almost constant at 0.2),
as is the vanishing of spin polarization for the next subband. Higher subbands
show an evolving mixture of HH/LH character in their spin-polarization profile.

A comparison of spin-polarization profiles (Fig.~\ref{fig:3}) and effective
$g$-factors (Table~\ref{tab:levels}) for each level yields a consistent picture.
States with mainly LH (HH) character, i.e., those having $\rho_1^2/\rho_0^2
\approx 1/5$ ($9/5$) can be associated with a definite spin polarization parallel
to the wire axis. Their corresponding $g^\ast$ is close to the expected value
$2\kappa$ ($6\kappa$). The lowest subband-edge state is an example for such
a polarized (LH) state. Other states exhibit a mixed spin-polarization profile, i.e.,
have both HH and LH polarizations present in different regions across the wire.
The associated $g$-factor values can vary widely.
Finally, states exist with vanishing hole-spin polarization over (most of)
the wire cross-section. By virtue of Eq.~(\ref{eq:InvRelate}), these states have a 
large spin-3/2 octupole moment. The second-lowest subband is an example for
this class of states, whose $g$-factor turns out to be (close to) zero. The
remarkable correspondence between local spin-polarization profiles and
$g$-factor values for hole-wire levels is another example for intriguing spin
physics emerging in nanostructures with strong spin-orbit coupling. The 
possibility to access individual properties of quasi-1D subband edges
in transport experiments~\cite{patel:prb:91,uz:prl:06} should enable the detailed
study of hole states with large and small spin polarizations and spin octupole
moments, respectively.

\textit{Principal caveats\/}. To gauge the relevance of our results for real
experiments, we need to discuss a few idealizations and approximations that are
implicit in our model. Most importantly, we neglected orbital magnetic-field
effects, band warping, and coupling to the conduction and split-off valence bands.
In addition, realistic wires will show deviations from perfect cylindrical symmetry
and usually have a softer than hard-wall confinement.

In principle, any physical property affecting HH-LH mixing will quantitatively
change $g^\ast$. It can be expected, however, that effects of band warping and
remote bands are small, as they turned out to be for 2D hole
systems~\cite{roland:prb:04}. Cross-sectional shape of a hole wire generally
affects its spin splitting~\cite{uz:pssc:06}. However, a recent numerical
study~\cite{kita:prb:06} found results for square CdTe wires that agree
closely with those presented in this work. A detailed comparison with
experiment will need to address orbital effects due to the typically
not-so-small magnetic fields used to measure $g^\ast$~\cite{patel:prb:91,
uz:prl:06,DMSappl}. Interpretation of data obtained in hole point
contacts~\cite{uz:prl:06} also requires a better understanding of the transition
region between 2D HH contacts and a constriction, and a proper treatment of
the strong quantum-well and soft lateral confinements.

\textit{Acknowledgments\/}. We thank P.~Brusheim, A.~F\"uhrer, M.~Governale,
A.R.~Hamilton, R.~Winkler, and H.Q.~Xu for useful discussions. DC acknowledges
support from the Massey University Research Fund. 


\begin{thebibliography}{31}
\bibitem{gFactDef}
Any two degenerate levels $E_{n m \pm}(0)$ are distinguished by quantum numbers
$\pm(m+3/2)$ of a new "total" angular momentum~\cite{sercel:prb:90} ${\hat F}_z =
{\hat J}_z-i\partial_\varphi$. Lifting of this degeneracy for finite $B$ implies that these 
doublets are equivalent to spin-1/2 degrees of freedom with $g$-factor given
by~(\ref{gStar}).
\end{thebibliography}

\end{document}